\documentclass[twocolumn]{aastex62}
\usepackage{comment}

\newcommand{\rulesep}{\unskip\ \vrule\ }

\received{\today}
\revised{\today}
\accepted{\today}
\submitjournal{ApJ}

\shorttitle{Stellar winds pump the heart of the Milky Way}
\shortauthors{Calder\'on et al.}
\begin{document}

\title{Stellar winds pump the heart of the Milky Way}

\correspondingauthor{Diego Calder\'on}
\email{diego.calderon@utf.mff.cuni.cz}

\author[0000-0002-9019-9951]{Diego Calder\'on}
\affiliation{Instituto de Astrof\'isica, Facultad de F\'isica, Pontificia Universidad Cat\'olica de Chile, 782-0436 Santiago, Chile}
\affiliation{Max Planck Institute for Extraterrestrial Physics, P.O. Box 1312, Giessenbachstr. 1, D-85741 Garching, Germany}
\affiliation{Institute of Theoretical Physics, Faculty of Mathematics and Physics, Charles University, 180 00 Prague, Czech Republic}

\author{Jorge Cuadra}
\affiliation{Instituto de Astrof\'isica, Facultad de F\'isica, Pontificia Universidad Cat\'olica de Chile, 782-0436 Santiago, Chile}

\author{Marc Schartmann}
\affiliation{Excellence Cluster Origins, Ludwig-Maximilians-Universit\"at M\"unchen, Boltzmannstr. 2, D-85748 Garching, Germany}
\affiliation{Universit\"atssternwarte der Ludwig-Maximilians-Universit\"at, Scheinerstr. 1, D-81679 M\"unchen, Germany}
\affiliation{Max Planck Institute for Extraterrestrial Physics, P.O. Box 1312, Giessenbachstr. 1, D-85741 Garching, Germany} 

\author{Andreas Burkert}
\affiliation{Universit\"atssternwarte der Ludwig-Maximilians-Universit\"at, Scheinerstr. 1, D-81679 M\"unchen, Germany}
\affiliation{Max Planck Institute for Extraterrestrial Physics, P.O. Box 1312, Giessenbachstr. 1, D-85741 Garching, Germany}
\affiliation{Excellence Cluster Origins, Ludwig-Maximilians-Universit\"at M\"unchen, Boltzmannstr. 2, D-85748 Garching, Germany} 

\author{Christopher M. P. Russell}
\affiliation{Instituto de Astrof\'isica, Facultad de F\'isica, Pontificia Universidad Cat\'olica de Chile, 782-0436 Santiago, Chile}



\begin{abstract} 
	The central super-massive black hole of the Milky Way, Sgr~A*, accretes at a very low rate making it a very underluminous galactic nucleus. 
	Despite the tens of Wolf-Rayet stars present within the inner parsec supplying ${\sim}10^{-3}\rm\ M_{\odot}\ yr^{-1}$ in stellar winds, only a negligible fraction of this material ($<10^{-4}$) ends up being accreted onto Sgr~A*. 
	The recent discovery of cold gas (${\sim}10^4\rm\ K$) in its vicinity raised questions about how such material could settle in the hostile (${\sim}10^7\rm\ K$) environment near Sgr~A*.
	In this work we show that the system of mass-losing stars blowing winds can naturally account for both the hot, inefficient accretion flow, as well as the formation of a cold disk-like structure. 
	We run hydrodynamical simulations using the grid-based code \textsc{Ramses} starting as early in the past as possible to observe the state of the system at the present time. 
	Our results show that the system reaches a quasi-steady state in about ${\sim}500\rm\ yr$ with material being captured at a rate of ${\sim}10^{-6}\rm\ M_{\odot}\ yr^{-1}$ at scales of ${\sim}10^{-4}\rm\ pc$, consistent with the observations and previous models. 
	However, on longer timescales ($\gtrsim3000\rm\ yr$) the material accumulates close to the black hole in the form of a disk. 
	Considering the duration of the Wolf-Rayet phase (${\sim}10^5\rm\ yr$), we conclude that this scenario likely has already happened, and could be responsible for the more active past of Sgr~A*, and/or its current outflow. 
	We argue that the hypothesis of the mass-losing stars being the main regulator of the activity of the black hole deserves further consideration. 
\end{abstract}



\section{Introduction} 
\label{sec:intro}

	 \cite{M19} recently detected cold gas (${\sim}10^4\rm\ K$) at short distances (${\sim}10^{-3}\rm\ pc$) from the Milky Way central super-massive black hole (SMBH), Sgr~A*. 
	The observation of a relatively wide (${\sim}2200\rm\ km\ s^{-1}$) double-peaked H30$\alpha$ recombination line suggests the presence of a disk-like structure made out of such material.   
	On the other hand, {\it Chandra} observations reveal that the innermost parsec of Sgr~A* is filled with hot (${\sim}10^{7}\rm\ K$) and diffuse plasma with a mean density of ${\sim}10\rm\ cm^{-3}$ \citep{B03,W13,R17}.  
	 This material is constantly being supplied by the powerful outflows of the tens of Wolf-Rayet (WR) stars inhabiting the region at $1$--$10\rm\ arcsec$ from Sgr~A* \citep{P06,Y14}\footnote{Given the mass and distance of Sgr~A* ($4.3\times10^6\rm\ M_{\odot}$ and $8.3\rm\ kpc$), $1\rm\ arcsec\approx 0.04\ pc \approx 10^5$ Schwarzschild radii.}. 
	Additionally, the G2 object \citep{G12}, as well as the bright knots in the IRS~13E cluster \citep{F10}, correspond to cold sources in this region that are observed in the Br-$\gamma$ recombination line. 
	How such cold gas structure could form in such a hot and hostile environment remains unexplained. 
	 
	 In total, the mass flow rate supplied by the WR stars is estimated to be ${\sim}10^{-3}\rm\ M_{\odot}\ yr^{-1}$ \citep{M07}. 
	 However, the detection of polarised emission at sub-mm wavelengths has constrained the accretion rate at hundreds of Schwarzschild radii (${\sim}10^{-5}\rm\ pc$) to be many orders of magnitude smaller \citep[$10^{-9}$--$10^{-7}\rm\ M_{\odot}\ yr^{-1}$;][]{Ma06,Ma07}. 
	 Although this is a very low accretion rate, it could have been different in the past.
	Since the early 90s, many authors have argued that Sgr~A* was more active in the past, specifically about hundreds of years ago \citep{S93,K96,P10}, and even as little as decades ago \citep{MB07}. 
	Such hypotheses rely on X-ray light echoes observations of molecular clouds located at hundreds of parsecs from the SMBH. 
	Based on long-exposure X-ray observations of the inner parsec, \cite{W13} ruled out a Bondi (steady state inflow) solution for the plasma distribution in the region. 
	Instead, the distribution showed agreement with an outflow solution, which supports the past activity hypothesis. 
	
\begin{figure}
		\centering
		\includegraphics[width=0.475\textwidth]{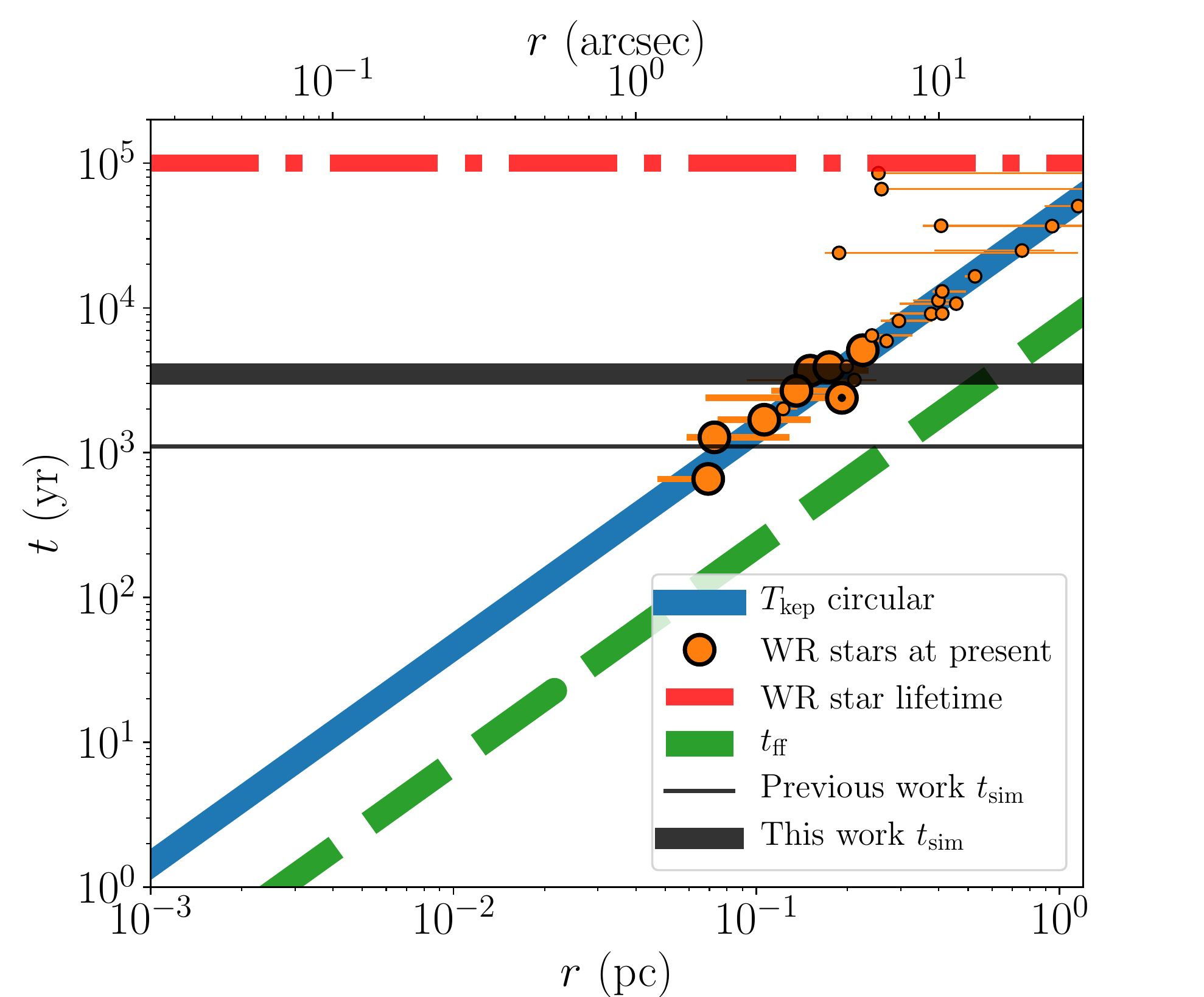}
		\label{fig:time}
		\caption{
		Timescales as a function of distance from Sgr~A*. 
		The solid blue line represents the circular Keplerian orbital period. 
		The dashed green line shows the free-fall timescale. 
		Orange circles denote the orbital periods of individual WR stars, with the horizontal orange segments indicating the separation change to Sgr~A* due to their orbital eccentricity. 
		Smaller symbols represent stars for which the eccentricities have been minimised when setting up their orbits (see Section~\ref{sec:setup}). 
		The circle with a black dot in its centre represents the star IRS~33E. 
		The horizontal dash-dotted red line marks the WR phase duration. 
		The horizontal solid thin and thick black lines indicate the simulation time of previous and current studies, respectively.}
	\end{figure}

	Numerical hydrodynamics models of the system of WR stars have been able to reproduce the mass inflow rate at the Bondi radius \citep[${\sim}10^{-6}\rm\ M_{\odot}\ yr^{-1}$;][]{C08,C15,R17}, and even deeper relatively well \citep{R18}. 
	These studies simulated the system starting $1100\rm\ yr$ in the past and evolved it to reach a quasi-steady state at the present. 
	Nevertheless, this timescale only manages to capture a single complete orbital period of the innermost WR star (see Figure~\ref{fig:time}). 
	Furthermore, it is approximately two orders of magnitude smaller than the typical duration of the WR phase in massive stars of $10$--$25\rm\ M_{\odot}$ \citep[${\sim}0.1\rm\ Myr$;][]{C07}. 
	In reality, the WR stellar system will exist for even longer as the stars do not necessarily evolve coordinately in and off the WR phase. 
	Thus, we should not draw strong conclusions of the long-term state of the system based on previous studies. 
	
	In this letter, we investigate, for the first time, the evolution of the system of mass-losing stars orbiting Sgr~A* for longer timescales. 
	Although we could not afford to simulate the system for timescales comparable to the typical duration of the WR phase (${\sim}0.1{\rm\ Myr}$), we improve the simulation time of previous works by a factor ${\sim}$three. 
	This work is presented as follows: Section~\ref{sec:setup} describes the numerical setup, initial conditions and assumptions. 
	Section~\ref{sec:results} presents the simulation and characterizes each phase of the evolution. 
	Finally, in Section~\ref{sec:disc} we discuss implications and present the conclusions.
		
	\begin{figure*}
		\centering
		\includegraphics[width=0.235\textwidth]{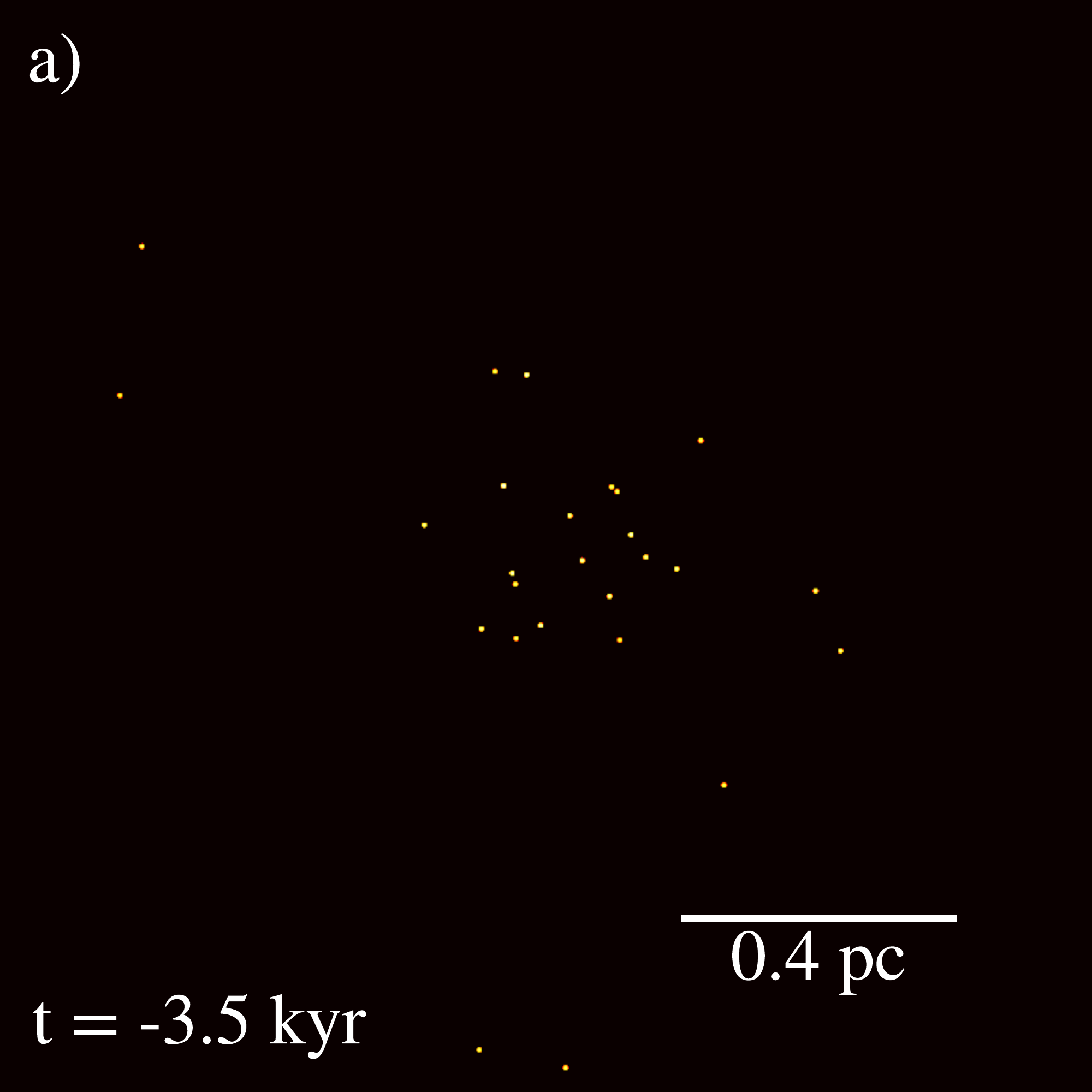}
		\hspace{-0.011\textwidth}
		\includegraphics[width=0.235\textwidth]{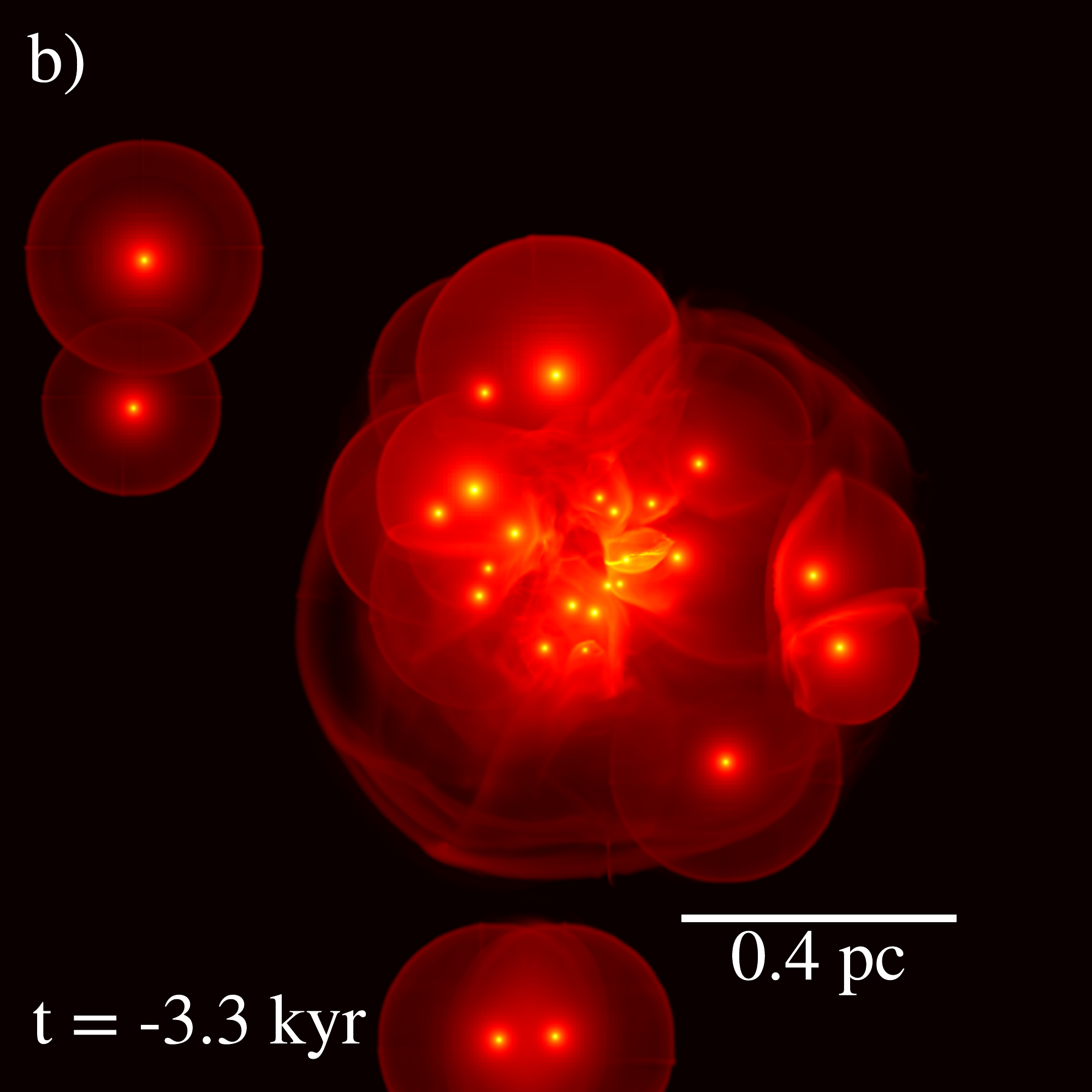}
		\hspace{-0.011\textwidth}
		\includegraphics[width=0.235\textwidth]{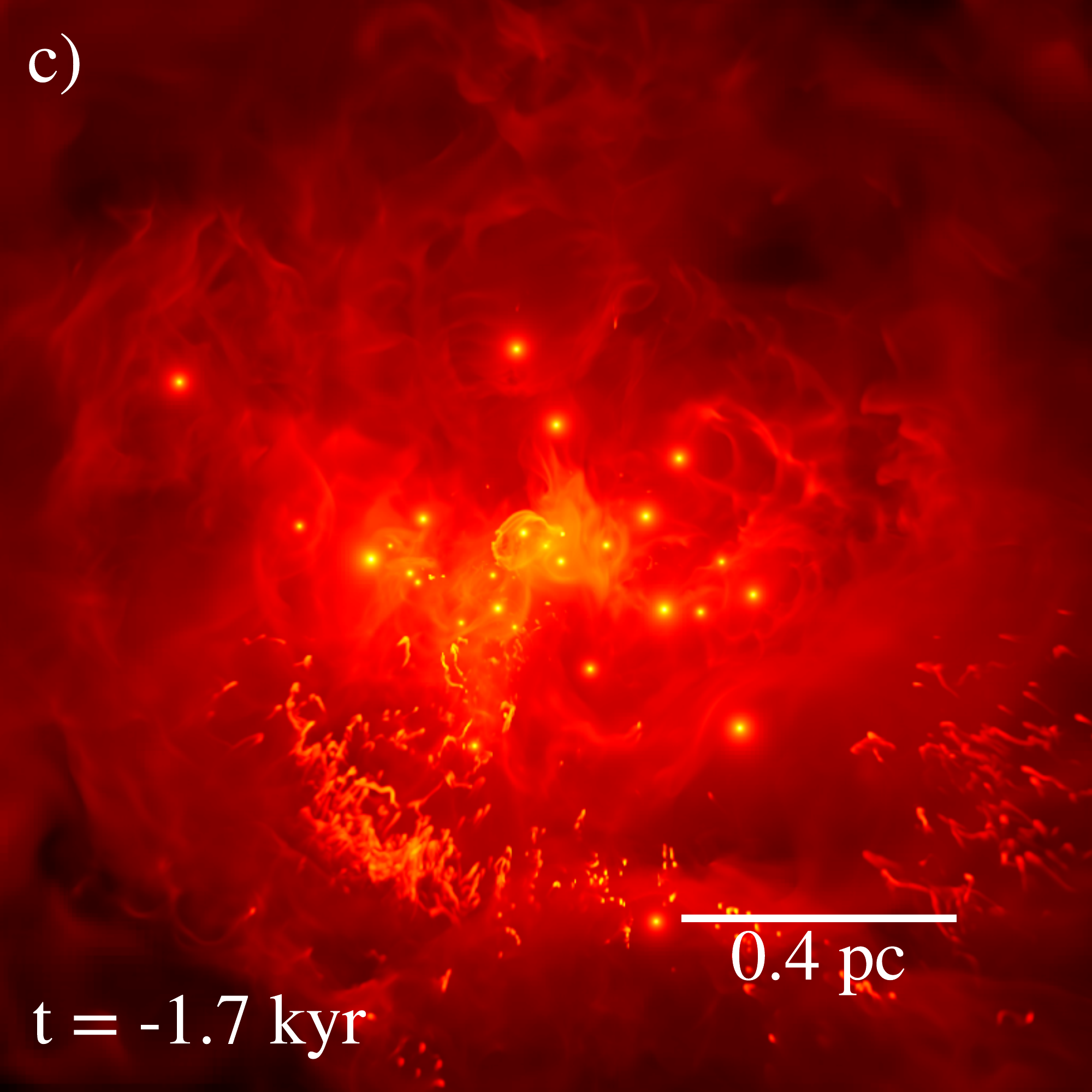}
		\hspace{+0.005\textwidth}
		\rulesep
		\hspace{+0.005\textwidth}
		\includegraphics[width=0.235\textwidth]{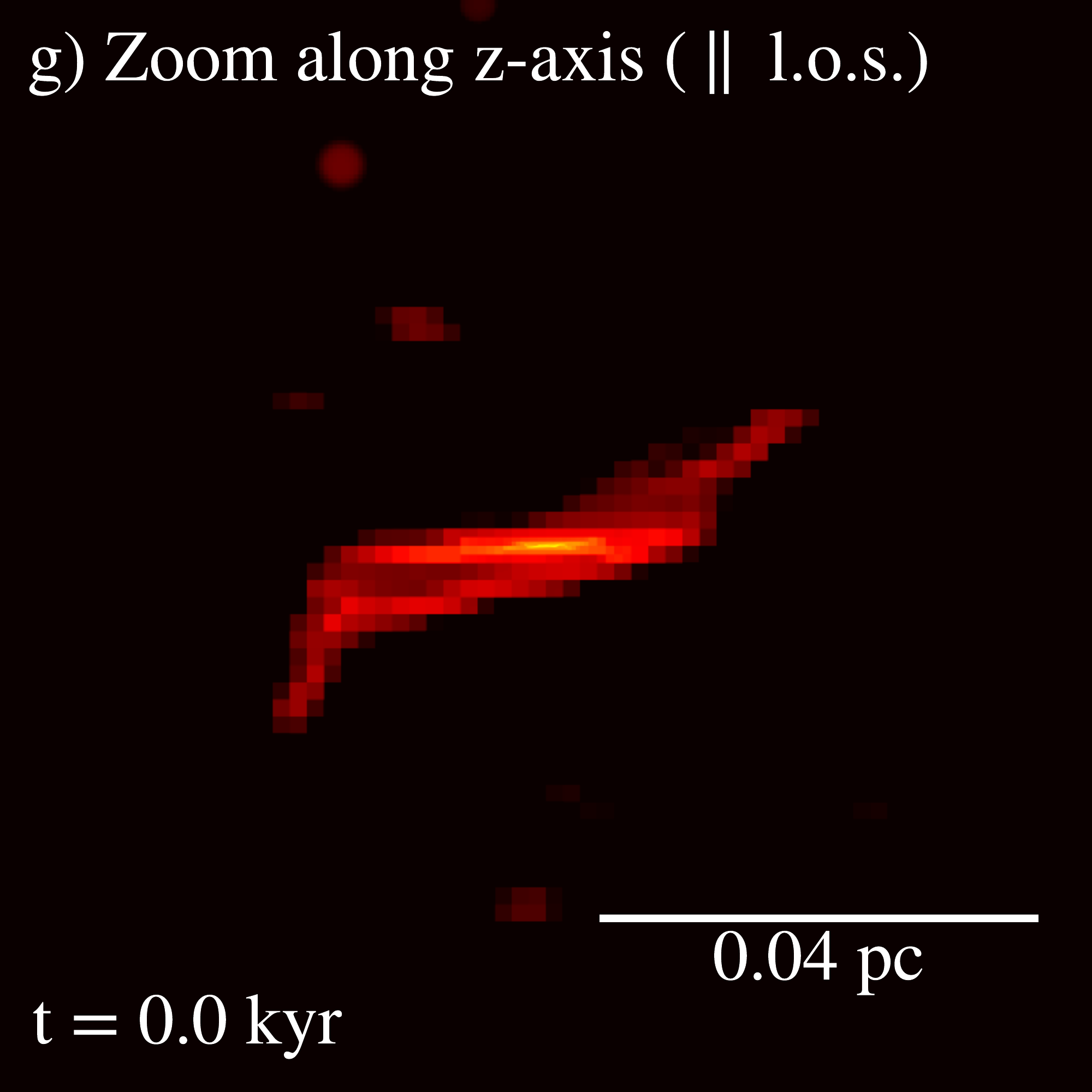}
		\includegraphics[width=0.235\textwidth]{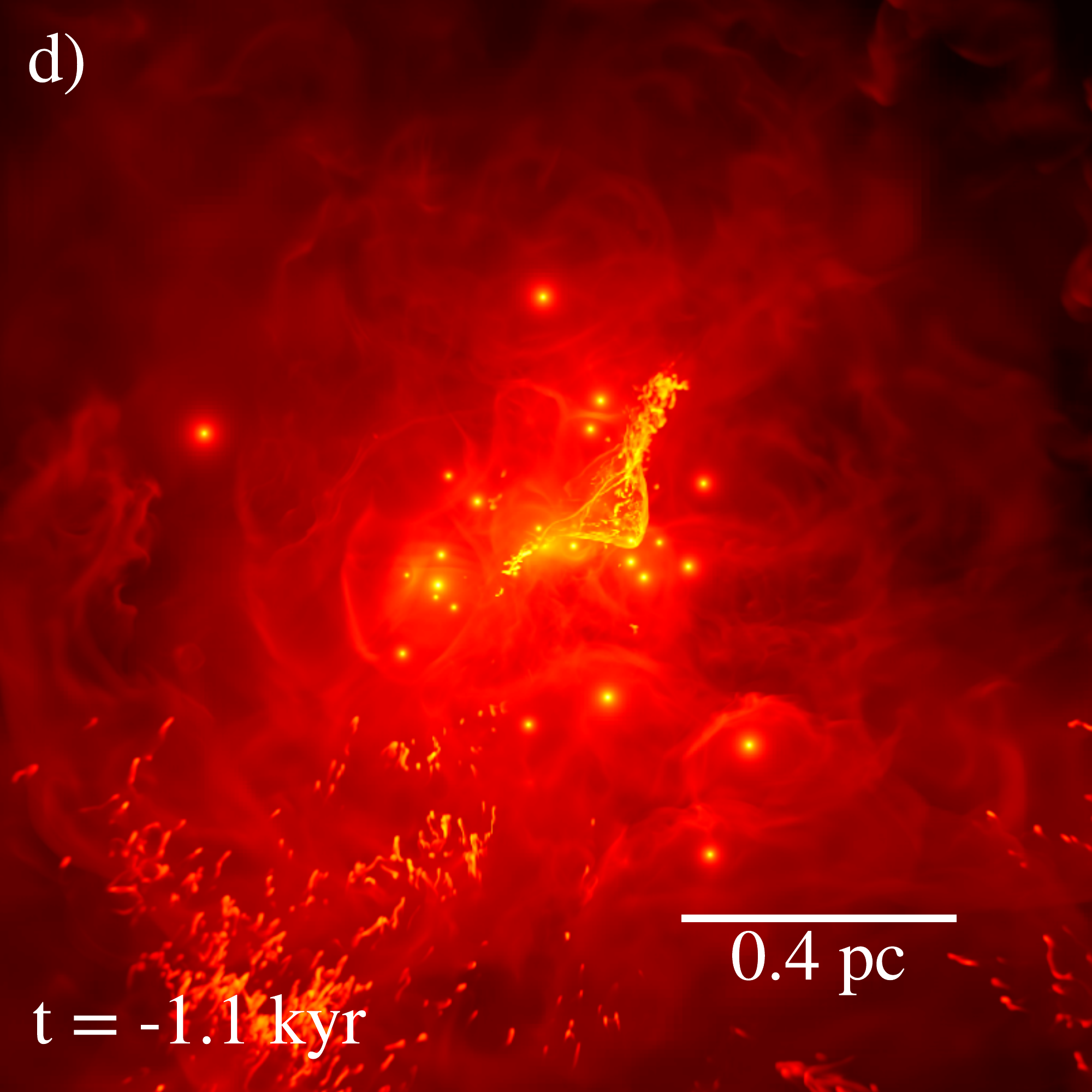}
		\hspace{-0.011\textwidth}
		\includegraphics[width=0.235\textwidth]{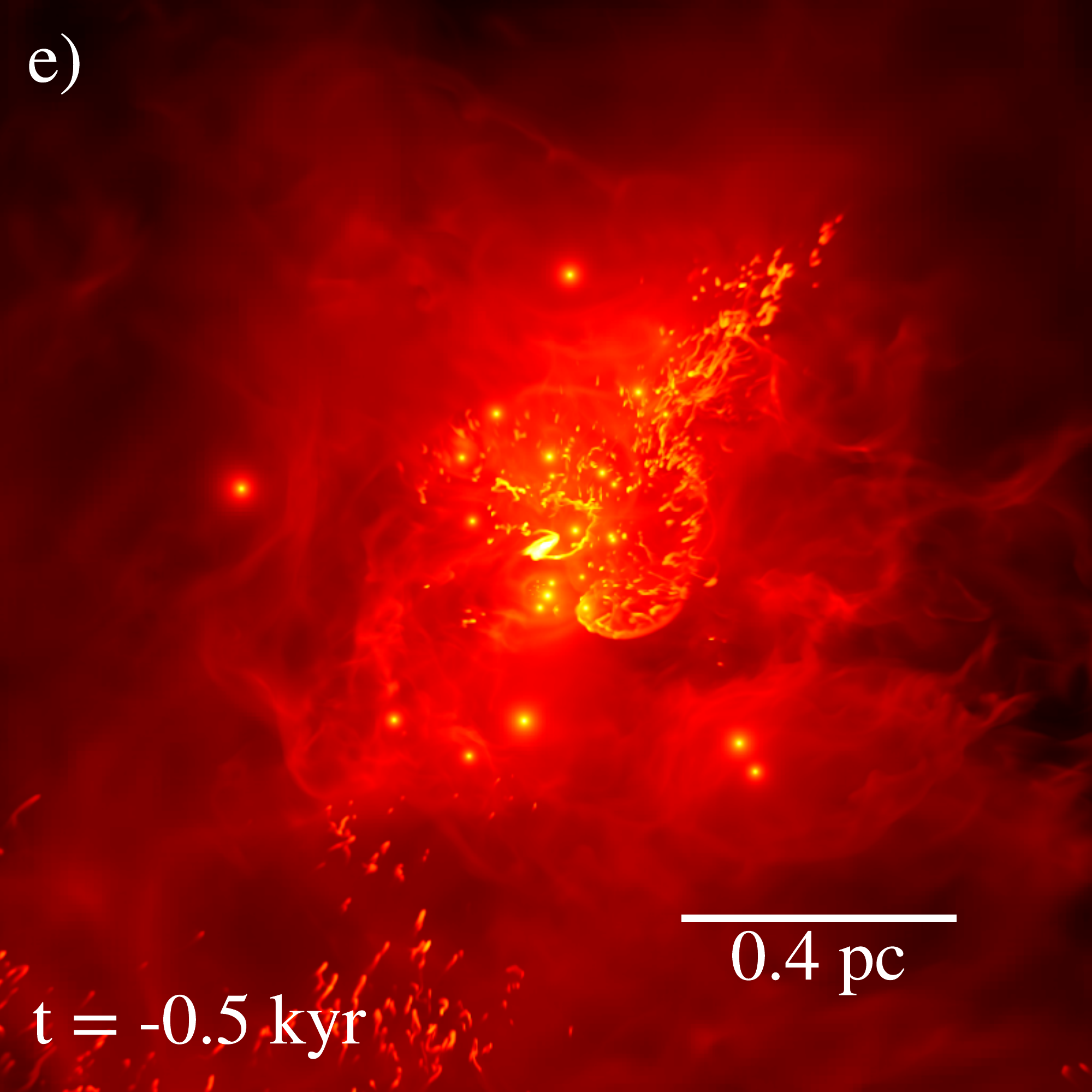}
		\hspace{-0.011\textwidth}
		\includegraphics[width=0.235\textwidth]{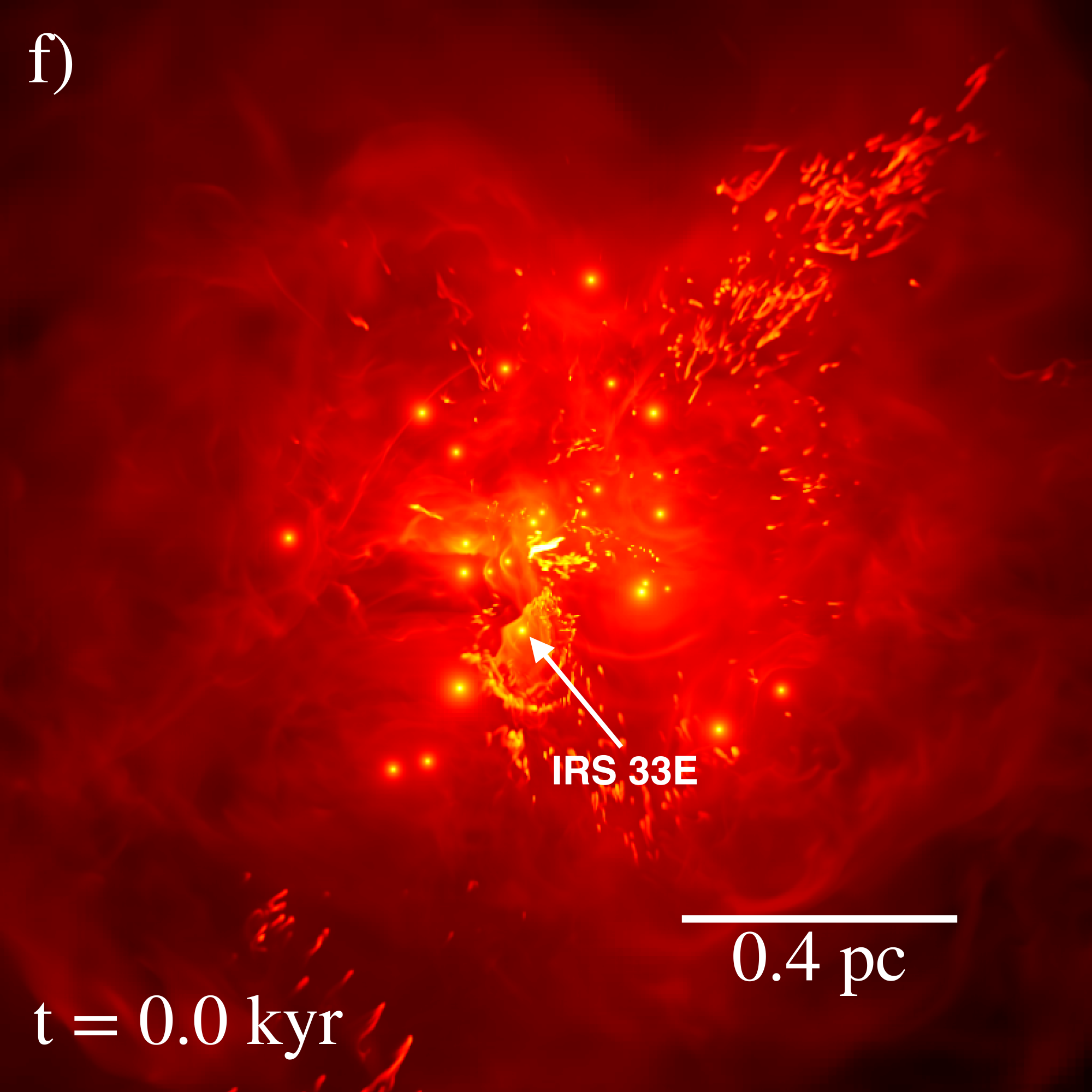}
		\hspace{+0.005\textwidth}
		\rulesep
		\hspace{+0.005\textwidth}
		\includegraphics[width=0.235\textwidth]{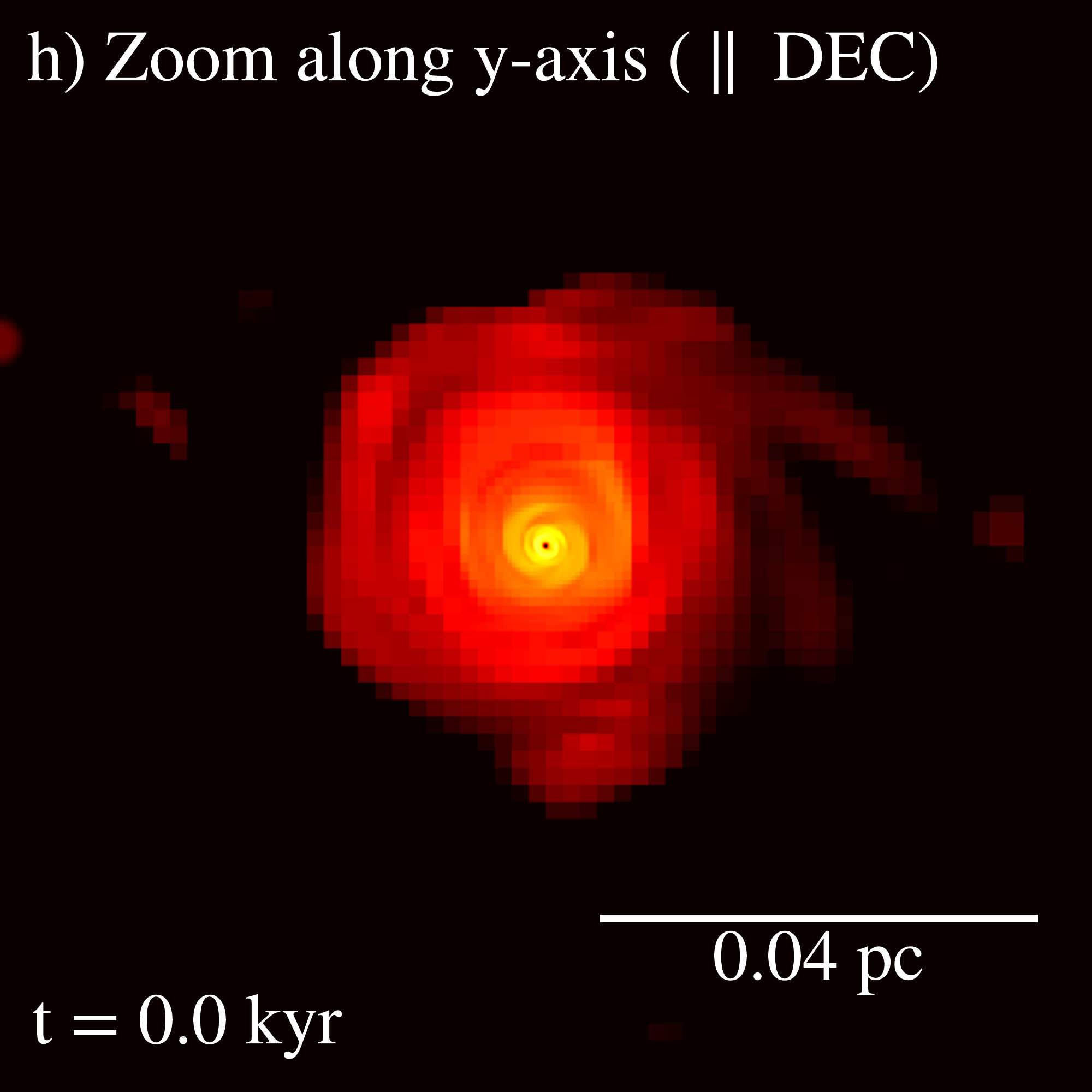}\\
		\hspace{0.001\textwidth}
		\includegraphics[width=0.65\textwidth]{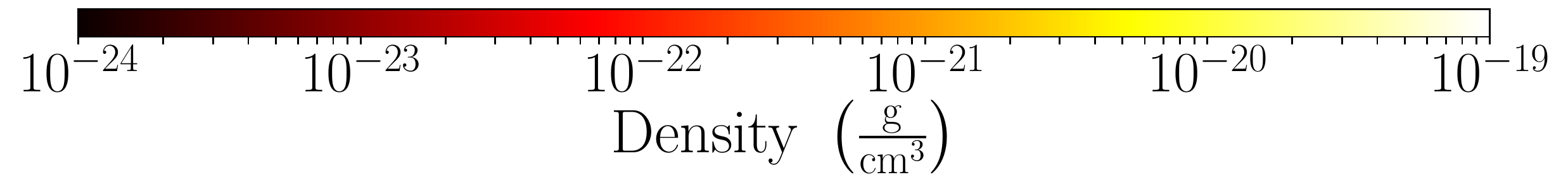}
		\hspace{0.03\textwidth}
		\rulesep
		\hspace{0.03\textwidth}
		\includegraphics[width=0.185\textwidth]{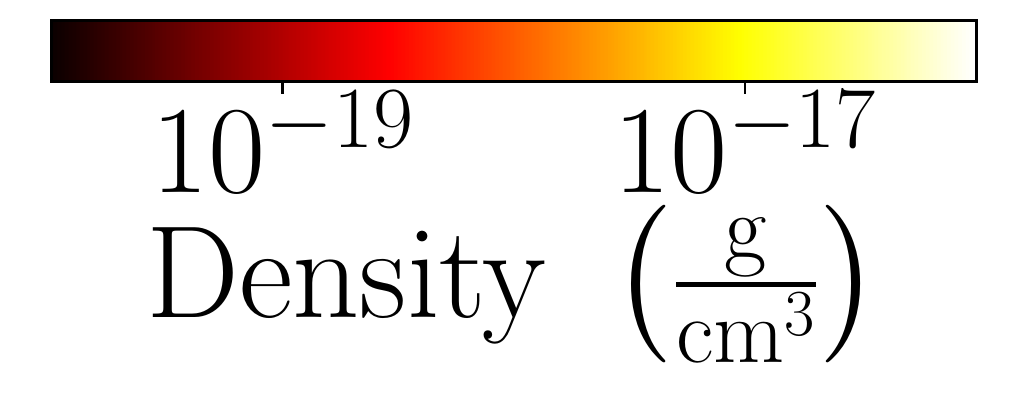}
		\caption{Complete evolution of the mass-losing stars orbiting Sgr~A*. 
		The panels a)-f) show projected density maps along the $z$-axis (parallel to the line-of-sight) weighted by density. 
		The horizontal and vertical axes are parallel to right ascension and declination, respectively. 
		Each panel shows the complete simulation domain at different simulation times. 
		Panels g) and f) are zoomed density projection maps at $t=0$ along the $z$- and $y$-axes, respectively, showing the disk-like structure around Sgr~A*.
		Notice the change in the scale of the colorbar.}
		\label{fig:evol}
	\end{figure*}

\section{Numerical setup}
\label{sec:setup}

	We developed 3D hydrodynamics simulations making use of the adaptive-mesh refinement (AMR) grid-based code \textsc{Ramses} \citep{T02}. 
	The code uses a second-order Godunov method to solve the equations of hydrodynamics. 
	We consider an adiabatic equation of state, plus optically thin radiative cooling with $Z=3Z_{\odot}$, where $Z_{\odot}$ is the Solar mass fraction in metals \citep[see][for a discussion]{C19}. 
	We use an exact Riemann solver together with a MonCen slope limiter \citep[e.g.,][]{T09}. 
	The simulation setup considers the system of WR stars blowing stellar winds into the medium while they move under the influence of the gravitational potential of Sgr~A* whose mass is $4.3\times10^6\rm\ M_{\odot}$ \citep{G17}. 
	For simplicity, we assume that the motion of the stars is completely determined by the gravity of the SMBH. 
	The stars describe Keplerian orbits, which have been constrained through decade-long observational monitoring \citep{P06,G17}. 
	For the stars whose orbits have not been completely determined, we used the most likely trajectories expected for the members of the ``clockwise disk" \citep{B06}, while the rest were calculated through minimizing their eccentricity \citep{C08}. 
	The stellar wind generation is simulated following the approach of \cite{L07}, and also used in \cite{C19}. 
	The properties of the stellar winds were taken from spectroscopic studies \citep{M07,C08}. 
	In order to avoid artificial accumulation of material close to the black hole we set an open boundary of radius $2\sqrt{3}\Delta x$, where $\Delta x$ is the size of the smallest cell. 
	This sphere in the domain is reset after each time step to low density at rest and low pressure \citep{R18}. 
	The domain of the simulation is a cubic box of side length $1.6\rm\ pc$ (${\sim}40\rm\ arcsec$) with outflow boundary conditions (zero gradients). 
	The coarse resolution corresponds to 64$^3$ cells, plus four extra refinement levels ($\Delta x\approx1.6\times10^{-3}\rm\ pc$). 
	Instead, the stellar wind generation regions allow five extra refinement levels ($\Delta x\approx 7.8\times10^{-4}\rm\ pc$), while the vicinity of the inner boundary allows eight extra refinement levels ($\Delta x\approx 9.8\times10^{-5}\rm\ pc$). 
	The initial position and velocity of the WR stars are determined extrapolating the state vectors to the past, so that the simulation evolves the system up to the present epoch \citep{C08,C15,R18}. 
	Most stars have periods of the order of ${\sim}10^4\rm\ yr$ and orbit around Sgr~A* at a distance of ${\sim}0.3\rm\ pc$ (see Figure~\ref{fig:time}). 
	Although at such distances the free-fall timescale (dashed green line) is of the order of hundreds of year, it is not easy for the stellar wind material to shed its angular momentum, so it corresponds to a lower limit of the actual infalling timescale. 
	
	In this work, the simulation starts from $3500\rm\ yr$ in the past in order to study whether it is possible for the system to reach, and maintain steady state or not\footnote{Ideally, we would like to start the simulation as early as possible but this incurs a very high computational cost: $3500\rm\ yr$ equates to ${\sim}100,000$ cpu hours.}. 
	As a control sample we run a model starting $1100\rm\ yr$ in the past, similar to previous works in the literature \citep{C08,C15,R18}. 
	The domain is initialized at very low density \hbox{$\rho_{\rm ISM}=10^{-24}\rm\ g\ cm^{-3}$} at rest \hbox{$\mathbf{u}=\mathbf{0}$}, and low pressure \hbox{$P_{\rm ISM}=\rho_{\rm ISM}c_{\rm s, f}^2\gamma^{-1}$}, where \hbox{$c_{\rm s, f}=10\rm\ km\ s^{-1}$} is the sound speed of the temperature floor (\hbox{$T\approx10^4\rm\ K$}), and $\gamma$ is the adiabatic index which is set to $5/3$ for an adiabatic gas. 
	This latter value was chosen, assuming that the strong ultraviolet radiation field of the young massive stars keeps the environment at such temperature.
	
	\begin{figure}
		\centering
		\includegraphics[width=0.475\textwidth]{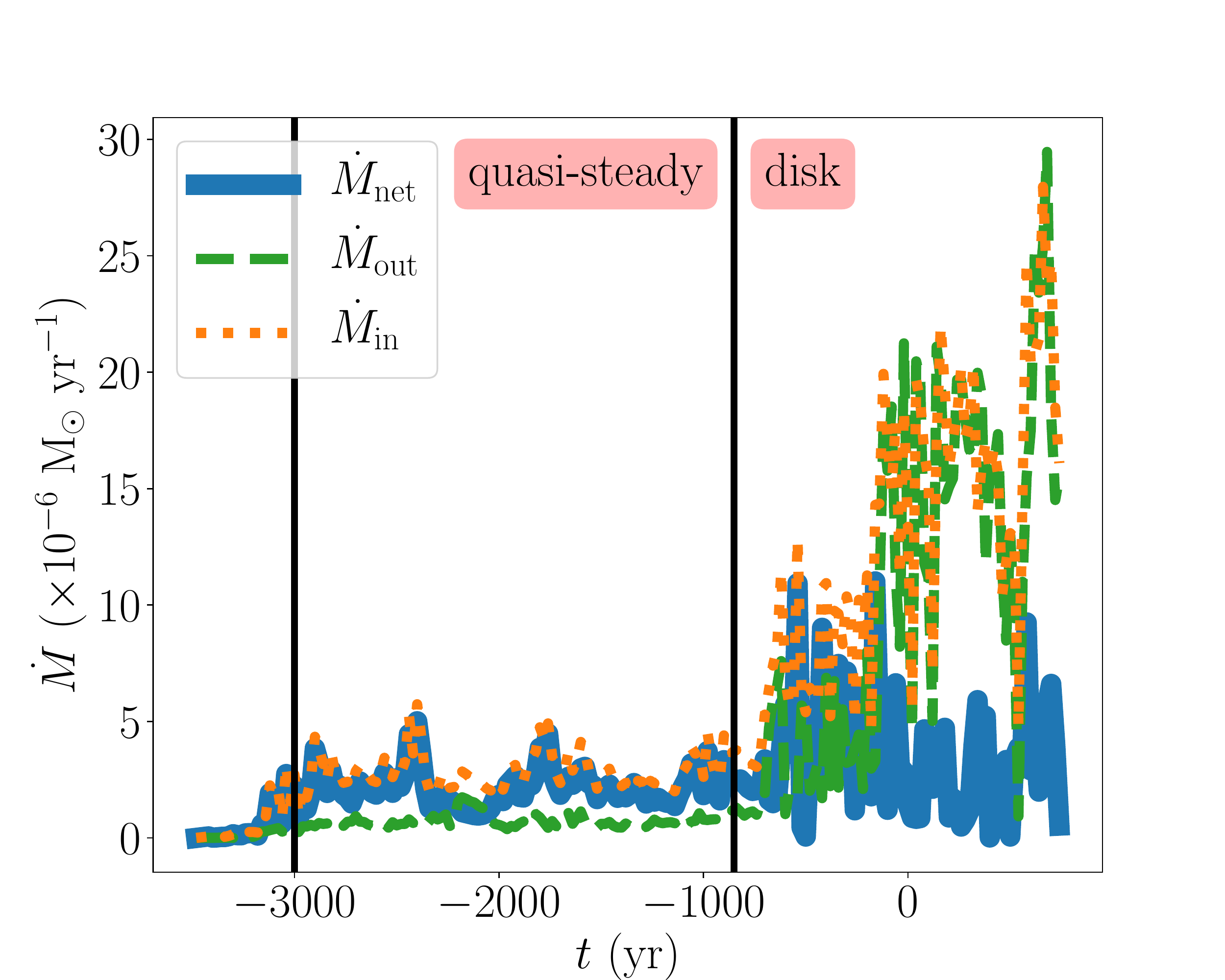}
		\caption{Mass flow rate across a sphere of radius $5~\times~10^{-4}\rm\ pc$ ($1.25\times10^{-2}\rm\ arcsec$). 
		The absolute value of the net mass flow rate is shown by the solid blue line. 
		The mass inflow and outflow rates are represented by the dotted orange and dashed green lines, respectively. 
		The vertical solid black lines (at $t\approx-3000\rm\ yr$ and $t\approx-800\rm\ yr$) divide the evolution into the transient, quasi-steady, and disk phases.}
		\label{fig:mdot}
	\end{figure}
	
	\begin{figure}
		\centering
		\includegraphics[width=0.475\textwidth]{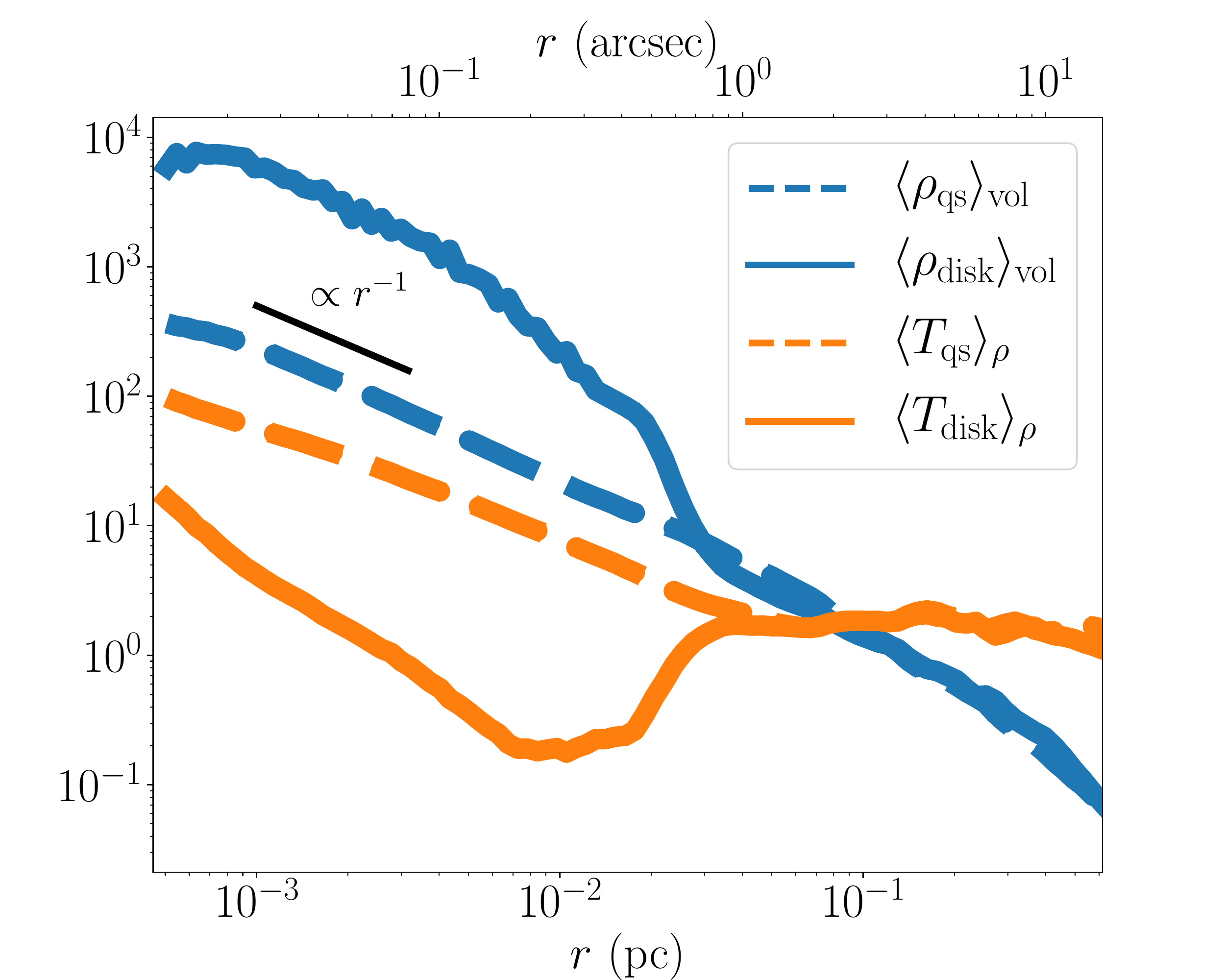}
		\caption{Density (blue lines) and temperature (orange lines) radial profiles in units of $10^{-22}\rm\ g\ cm^{-3}$ and $10^7\rm\ K$, respectively. 
		The subscripts ``qs" and ``disk" refer to time-averaged quantities during the quasi-steady state (dashed lines) and the disk phase (solid lines), respectively.}
		\label{fig:prof}
	\end{figure}
	
	\begin{figure*}
		\centering
		\includegraphics[width=0.6\textwidth]{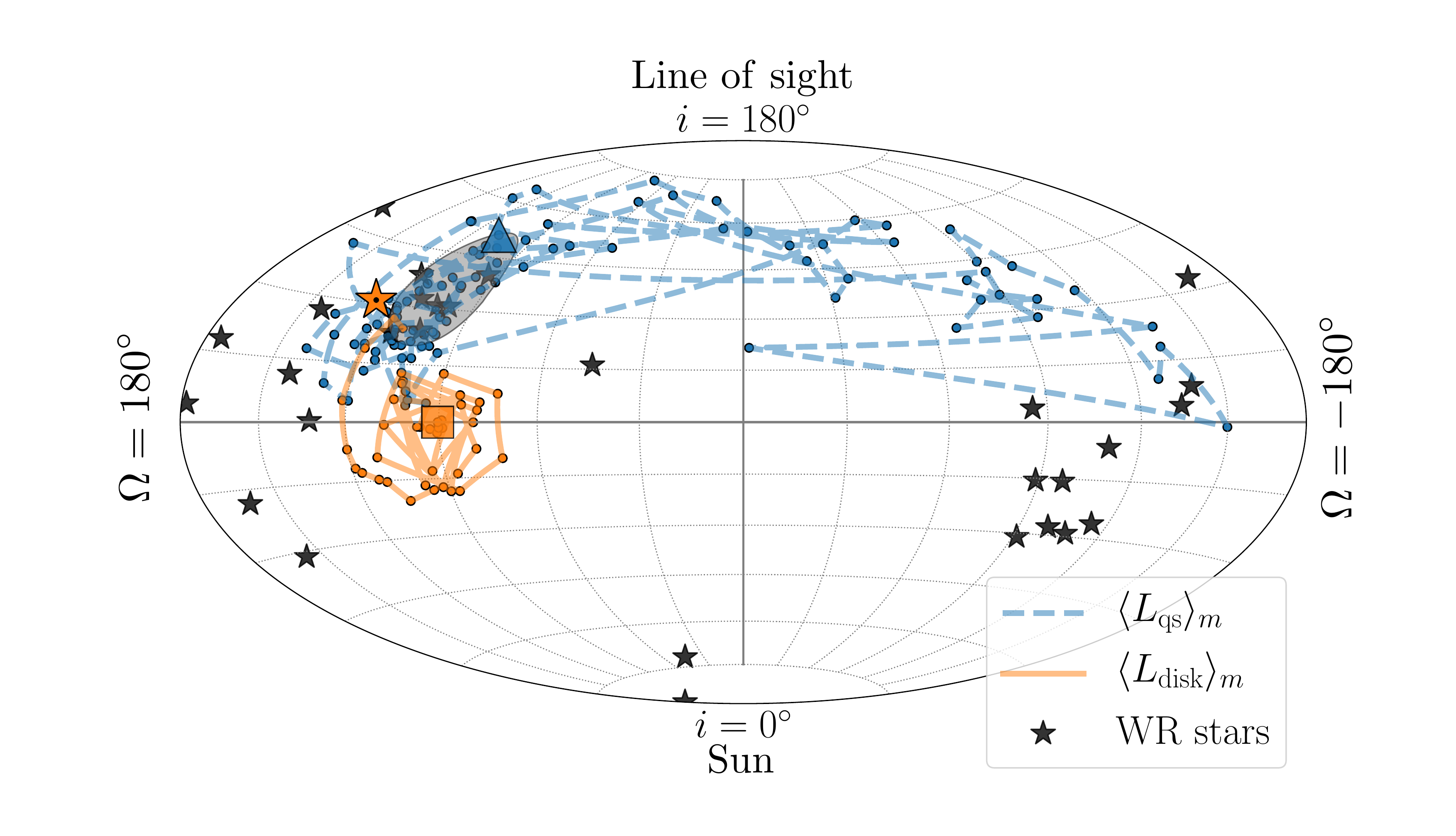}
		\caption{
		Orientation of the angular momentum of the gas enclosed in a sphere of radius $0.01\rm\ pc$ ($0.25\rm\ arcsec$), and the WR stars'. 
		The vertical dimension represents inclination $i$, while the horizontal dimensions stand for the longitude of the ascending node $\Omega$. 
		Thus, a face-on star orbiting clockwise on the sky would be at the north pole of the graph.  
		The dashed blue and solid orange lines represent the quasi-steady and disk phases, respectively. 
		Each dot corresponds to the analysis of a single snapshot. 
		The triangle and square show the initial and final state, respectively. 
		The black stars represent the angular momentum direction of the WR stars, highlighting IRS~33E as an orange star with a black dot, instead. 
		The grey shaded region corresponds to the direction of the clockwise disk at $(104^{\circ},126^{\circ})$ with a thickness of $16^{\circ}$ \citep{Y14}.}
		\label{fig:hammer}
	\end{figure*}

\section{Hydrodynamics: evolution phases}
\label{sec:results}

	Figure~\ref{fig:evol} shows density maps illustrating the time evolution of the simulation. 
	Each map corresponds to a density projection of the full domain along the $z$-axis, which is parallel to the line-of-sight, weighted by density (i.e., ${\int}{\rho^2}{dz}{/}{\int}{\rho}{dz}$)\footnote{This quantity helps to highlight the dense gas, which corresponds to the cold material which is most likely to contribute to the disk formation and the most refined regions of the domain.}. 
	The horizontal and vertical axes are parallel to right ascension and declination, respectively. 
	Panel a) shows the initial condition of the simulation at $t={-}3500\rm\ yr$. 
	Once the simulation starts, each star blows a wind, which develops a shock that at some point ends up colliding with the others (panel b). 
	Such collisions create dense (${\sim}10^{-22}\rm\ g\ cm^{-3}$), hot material (${\sim}10^7\rm\ K$) that fills the entire domain relatively quickly (${\sim}500\rm\ yr$).
	Notice how complex structures develop in the gas, e.g., the interaction of stellar winds, instabilities, dense clumps, and bow shocks forming due to the motion of the stars through the medium. 
	This depends on cooling being efficient, which is sensitive to the stellar wind abundances. 
	Fortunately, such abundances have been constrained through spectroscopic studies relatively well \cite[see][for a more detailed justification]{C19}.
	In panel c) the system has already attained what appears to be a quasi-steady state like in previous works in which the simulations end after a time comparable to the evolution time of panel d) \citep{C08,R18}. 
	However, in panel g) it is possible to observe an accumulation of material at the center of the domain. 
	Visually, this material seems to settle into a disk-like structure around Sgr~A*, which is confirmed analyzing the central region in detail (see panels g and h). 
	This material seems to mainly arise from the strong bow shock generated by the star IRS~33E traveling through the dense medium (panels~d-f). 
	Based on the appearance of the system we divide its evolution in three phases: the transient, quasi-steady, and disk phases. 
	
	Figure~\ref{fig:mdot} shows the mass flow rate across a sphere of radius ${\sim}5\times10^{-4}\rm\ pc$ ($1.25\times10^{-2}\rm\ arcsec$) as a function of time, highlighting the transition between phases. 
	This length scale is about two orders of magnitude smaller than the radius of the disk.
	In the transient phase the mass inflow rate increases up to ${\sim}10^{-6}\rm\ M_{\odot}\ yr^{-1}$ roughly at \hbox{$t\approx{-}3000\rm\ yr$}. 
	Then, in the quasi-steady phase the net mass flow rate $\dot{M}_{\rm net}$ is variable but permanently inflowing and around the same order of magnitude. 
	At $t\approx{-}800\rm\ yr$ the system enters into the disk phase, whereby the formation and presence of the disk produces significant changes in both the inflow and outflow mass flow rates. 
	In general, the net flow remains variable, but now its amplitude is about one order of magnitude larger due to the enhancement of the inflow and outflow rates. 
	Notice that this behavior is observed even at $t>0$.
	
	Figure~\ref{fig:prof} shows the time-averaged density (blue lines) and temperature (orange lines) radial profiles. 
	The averages were calculated over the quasi-steady (dashed lines) and the disk (solid lines) phases. 
	Notice that in the quasi-steady phase both profiles decay with $r^{-1}$ in the innermost region ($r\lesssim0.03\rm\ pc$), resembling the simulation by \cite{R18}. 
	Although not shown here, the control run (started at $t=-1100\rm\ yr$) displays the same behavior at the present time. 
	In the disk phase, the profiles deviate significantly from their quasi-steady phase shapes. 
	The disk formation increases the density by about an order of magnitude, which extends up to ${\sim}0.01\rm\ pc$ from Sgr~A*. 
	On the contrary, the temperature profile decreases by an order of magnitude but, at the same time, becomes slightly steeper. 
	Analyzing the disk structure at the end of the simulation, we observe that the disk has a mass of ${\sim}5\times10^{-3}\rm\ M_{\odot}$, with a temperature of ${\sim}10^4\rm\ K$. 
	We ran the model for an extra $1000\rm\ yr$ into the future and found that the disk is not destroyed, instead its mass increases at an average rate of ${\sim}10^{-6}\rm\ M_{\odot}\rm\ yr$.
	
	Figure~\ref{fig:hammer} is a Hammer projection as seen from the Sgr~A* location that shows the evolution of the direction of the averaged angular momentum of the gas enclosed in a sphere of radius $0.01\rm\ pc$ ($0.25\rm\ arcsec$) in the quasi-steady (dashed blue line), and disk (solid orange line) phases. 
	The angles $\Omega$ and $i$ correspond to the longitude of the ascending node and inclination, respectively. 
	The earliest time is represented with a triangle marker, while the latest with square marker. 
	As a reference, we include the angular momentum direction of the WR stars (black stars), as well as of the stellar clockwise disk at $\Omega=104^{\circ}$ and $i=126^{\circ}$ \citep{Y14}. 
	Notice that in the quasi-steady phase the angular momentum direction is highly variable due to the stochastic accretion of material coming from different stars that have a close passage from Sgr~A*. 
	However, most of the time the angular momentum direction is consistent with the stars, whose orbits are located near the clockwise disk.
	This is expected given that these stars have relatively slower winds (${\sim}600\rm\ km\ s^{-1}$) and orbit closer to the black hole ($\lesssim0.3\rm\ pc$), such that their wind angular momentum is smaller \citep{C08,R18}. 
	In the disk phase, initially the angular momentum of the gas is aligned with the stellar disk, and more specifically with IRS~33E, but ultimately precesses around $(\Omega,i)\approx(90^{\circ},0^{\circ}$). 
	Notice that this precession could be an undesired effect due to the Cartesian grid.  

\section{Discussion}
\label{sec:disc}

	We have shown that, if modeled for a long enough timescale (${\gtrsim}3000\rm\ yr$), the natural outcome of the evolution of the WR stellar system is a disk-like structure around Sgr~A*.
	The crucial conditions for this to occur are the interaction between the wind of one particular star, IRS~33E, and the dense medium, together with modelling the system at least a complete orbit with such a dense medium built. 
	Although the mass loss rate of the star is average for a WR type (${\sim}1.6\times10^{-5}\rm\ M_{\odot}\ yr^{-1}$), its wind speed is the slowest among all the stars \citep[$450\rm\ km\ s^{-1}$;][]{M07}. 
	Thus, the ram pressure of its wind is weak, which forces it to be denser at the stagnation point. 
	As a result, the shocked material radiates most of its thermal energy almost instantaneously \citep[see Section 4.1 in][for a discussion]{C16}. 
	This can be seen in the form of a dense, cold bow shock in front of the star while orbiting around Sgr~A* (see panel d of Figure~\ref{fig:evol}). 
	Figure~\ref{fig:time} shows that the pericenter and apocenter distances of the star are $0.069\rm\ pc$ and $0.213\rm\ pc$ from the SMBH, respectively. 
	Therefore, the star oscillates between the regimes where the net mass flow to Sgr~A* is ${\sim}0$ and where the outflow dominates \citep[see Figure~12 of][]{R18}. 
	In this scenario, the star feeds the nucleus during every pericenter passage of its ${\sim}2000\rm\ yr$ orbit (see Figure~\ref{fig:time}).  
	However, this occurs thanks to the presence of a dense medium that interacts with the wind. 
	In contrast, the control run does not show this behavior likely due to limited simulation time once the dense environment is established. 

	Nonetheless, there are a couple of caveats we have to bear in mind when interpreting this result. 
	High-resolution simulations of idealized stellar wind collisions have shown that the dense clumps formed are neither massive nor large \citep[$\lesssim10^{-3}\rm\ M_{\oplus}$;][]{C19}. 
	The parameters of such models were largely motivated by the WR stars in the Galactic Center. 
	In principle, this would suggest that the clumpy structure observed in the larger-scale simulation presented here might not be resolved, being smaller and less massive in reality. 
	That being said, it is not straightforward to extrapolate the earlier result to our current configuration, as the mechanism creating clumps here is different. 
	Being so, the disruption of a dense stellar wind bow shock in this environment is a promising channel for clump formation, and possibly G2-like objects \citep{B12}, unlike {\em colliding} winds \citep{C16,C18,C19}. 
	
	In order to check potential numerical effects on the result we ran a couple of extra tests. 
	Firstly, we used a MinMod flux limiter, which resulted in the same observed behavior, or even better as the disk did not precess and remained aligned with the clockwise disk. 
	Nevertheless, one of the caveats with this approach is that the winds are not spherically symmetric for reasonable computational costs.	
	Secondly, we decreased the density threshold of the refinement criterion, which results in a more uniform grid. 
	The same behavior was observed but the disk formed slightly later. 
	
	Overall, the accumulation of material around Sgr~A* takes place on timescales of thousands of years, which leads to an enhancement in both the mass inflow and outflow rates (see Figure~\ref{fig:mdot}). 
	Despite the higher accretion, once the disk forms, Sgr~A* is still in the so-called radiatively inefficient accretion flow (RIAF) regime.
	In this context, as the material is very hot, it can overheat and then eject a big fraction of mass and energy in the form of a strong outflow \citep{B99,Be12}. 
	The disk seen in the simulation could then have produced the inferred larger luminosity for Sgr~A* \citep{P10}, and/or triggered the reported outflow \citep{W13,D19}. 
	In that case, it is not clear whether the disk still exists or if it was destroyed by such an outflow. 
	But the cold gas reported by \cite{M19}, if confirmed as a disk, could correspond to the one from our models.
	In fact, in order to match the H30$\alpha$ and Br-$\gamma$ observational constraints, we require an amplification factor of ${\sim}30$, consistent with the masing factor ${\sim}80$ reported by \cite{M19}.
	However, considering the expected interaction of G2 with this disk, the density we find is an order of magnitude higher than the one derived by \cite{G19} from the cloud slow-down.
	
	To conclude, we speculatively propose that the stellar winds alone can be responsible for much of the phenomenology observed and/or inferred in the central arcsecond during the last millenium: the variable accretion and luminosity, a cold disc, and an outflow. 
	If so, there is no need to invoke ``external'' factors such as infalling gas from Sgr A West (a.k.a., the \textit{minispiral}), a tidal disruption event and/or a supernova.
	
\acknowledgments
\scriptsize
	We thank the anonymous referee for useful comments and suggestions to improve this article.
	DC and JC acknowledge the kind hospitality of the Max Planck Institute for Extraterrestrial Physics as well as funding from the Max Planck Society through a ``Partner Group'' grant. 
	We thank F. E. Bauer and J. Dexter for useful discussions and suggestions for improving this work.  
	The authors acknowledge support from \textsc{conicyt} project Basal AFB-170002. 
	DC is supported by \textsc{conicyt}-\textsc{pcha}/Doctorado Nacional (2015-21151574). 
	CMPR is supported by \textsc{fondecyt} grant 3170870. 
	Numerical simulations were run on the high-performance computing system \textsc{cobra} of the Max Planck Computing and Data Facility. 
	Data analysis was carried out making use of the \textsc{python} package \textsc{yt} \citep{T11}.

\end{document}